# Correlated Excitonic Signatures in a Nanoscale van der Waals Antiferromagnet

Vigneshwaran Chandrasekaran,[1*] Christopher R. DeLaney,[1] David Parobek,[1] Christopher A. Lane,[2] Jian-Xin Zhu,[1,2] Xiangzhi Li,[1] Huan Zhao,[1] Cong Tai Trinh,[1] Marshall A. Campbell,[1] Andrew C. Jones,[1] Matthew M. Schneider,[3] John Watt,[1] Michael T. Pettes,[1] Sergei A. Ivanov,[1] Andrei Piryatinski,[2] David H. Dunlap,[4*] Han Htoon[1*]

[1] Center for Integrated Nanotechnologies, Materials Physics and Applications Division, Los Alamos National Laboratory, Los Alamos, New Mexico 87545, United States

[2] Theoretical Division, Los Alamos National Laboratory, Los Alamos, New Mexico 87545, United States

[3] Materials Science in Radiation and Dynamics Extremes, Materials Science and Technology Division, Los Alamos National Laboratory, Los Alamos, New Mexico 87545, United States

[4] Department of Physics and Astronomy, University of New Mexico, Albuquerque, New Mexico 87131, United States

**Abstract**

**Composite quasi-particles with emergent functionalities in spintronic and quantum information science can be realized in correlated materials due to entangled charge, spin, orbital, and lattice degrees of freedom. Here we show that by reducing the lateral dimension of correlated antiferromagnet $NiPS_3$ flakes to tens of nanometers, we can switch-off the bulk spin-orbit entangled exciton in the near-infrared (1.47 eV) and activate visible-range (1.8 – 2.2 eV) transitions with charge-transfer character. These ultra-sharp lines (<120 μeV at 4.2 K) share the spin-correlated nature of the bulk exciton by displaying a Néel temperature dependent linear polarization. Furthermore, exciton photoluminescence lineshape analysis reveals a polaronic character via coupling with at-least 3 phonon modes and a comb-like Stark effect through discretization of charges in each layer. These findings augment the knowledge on the many-body nature of excitonic quasi-particles in correlated antiferromagnets and also establish the nanoscale platform as promising for maturing integrated magneto-optic devices.**

Introduction

Excitons – quasi-particle bound states of electron-hole pairs – are investigated extensively as they serve as the foundation of almost all modern photonic technologies.[1] The efforts of the past decades focused on uncorrelated excitons in conventional semiconductors where quasi-particles weakly interact with one another. Currently, correlated excitons are generating significant interest due to their intrinsic many-body interactions entangling charge, spin, orbital, and lattice degrees of freedoms.[2-4] A wide breadth of emergent quantum phenomena beyond the physics of conventional semiconductors, such as excitonic insulators,[5] exciton condensates,[6] and exciton Hall effects,[7] have been recently reported, while charge fractionalization[8] and exciton-mediated superconductivity[9] are still elusive. Among a few classes of materials capable of supporting such exciting phenomena, *e.g.* transition metal dichalcogenides,[10] topological insulators,[11] and twisted bilayer graphene,[12] the correlated van der Waals magnets have risen to the forefront due to the coexistence of long-range magnetic correlations, strong electron-phonon coupling, and a robust excitonic response all under two-dimensional (2D) confinement.[13-15]

A member of this material family, $NiPS_3$, is a charge-transfer antiferromagnetic (AFM) insulator similar to the hole-doped high-$T_c$ cuprates.[16] Owing to strong Ni-*d* and S-*p* hybridization in the valence (conduction) bands,[17] a co-existence of *d-d* and charge-transfer electron-hole pairs is expected.[18] Concomitantly, several optical transitions from UV to NIR are reported in absorption,[19] reflectance,[20] and photo-conductivity studies.[16] While the strongest peak is assigned to a charge-transfer state around 2.2 eV that persists even up to room temperature,[16] photoluminescence (PL) spectroscopy of bulk samples reveals a correlated exciton at lower energies (1.47 eV) that arises intrinsically from the many-body states of the Zhang-Rice singlet upon cooling below the Néel transition temperature $T_N$ (150 K).[21] This exciton is assigned to a spin-orbit entangled state (SO-X) displaying an ultra-narrow linewidth,[21] phonon-bound states,[20] spin-correlated behavior,[22] thickness-dependent PL,[21] polariton formation,[23] and can be controlled using light-pumping[24] and external magnetic fields.[22] However, to date, whether or not reduction of the lateral dimension influences these fascinating phenomena has not been explored. This approach would not only benefit from the scalable top-down and bottom-up fabrication approaches, but could also advance the fundamental understanding of excitons in nanoscale correlated insulators in parallel with excitons in nanoscale semiconductors.[25] Aiming to address this question, we have conducted single nanostructure optical spectroscopy studies on individual nanoflakes (NF) of $NiPS_3$.

**Results**

We synthesize NFs of $NiPS_3$ in solution form by means of metathesis reaction (see Methods for details). Their crystalline quality is verified by HR-TEM of an entire NF shown in Fig. 1a (also see Supplementary Information S.1 for XRD). The lateral sizes of our NFs are in the range of 20-100 nm with thicknesses ranging from 1-10 nm obtained from scanning probe images (Fig. 1b). While the sharp many-body excitonic transition (SO-X at 1.47 eV) reported for bulk $NiPS_3$ can be observed on flakes with lateral sizes as small as 1 µm (see Supplementary Information S.2) and 3-layer thickness,[21] this emission completely disappears for the NFs. Instead, various individual NFs emit PL peaks between 1.8 eV and 2.2 eV in the visible spectral range (Fig. 1c). At 4.2 K, the main PL peak of our NF is characterized by a spectral resolution-limited linewidth of 120 µeV (Fig. 1c), which is three times narrower than that reported for SO-X bulk exciton.[22] A time resolved PL study (Fig. 1d) reveals a lifetime of ~3 ns, which is more than two orders of magnitude longer than ~10 ps reported for the bulk SO-X peak.[20,22] The flake-to-flake variation of emission energies with narrow linewidths and long PL decay times are similar to excitons confined in quantum dots and defects. However, the 20-100 nm size of our NFs is not sufficient to provide a strong quantum confinement to induce 300-700 meV blueshift from 1.47 eV bulk peak since the bulk exciton is reported to have a high binding energy of over several hundred meV with a small Bohr radius of ~0.6 nm.[23] Second-order autocorrelation measurements also yield Poissonian photon-statistics (Fig. 1e) instead of the photon-antibunching typically observed in quantum dots and defects. Our pump-dependent PL study (Fig. 1f and Supplementary Information S.3) further shows that the intensity of the sharp emission peaks increases linearly over 3 orders of magnitude variation in pump power, in contrast to a saturation expected for a fixed number of localized excitons.

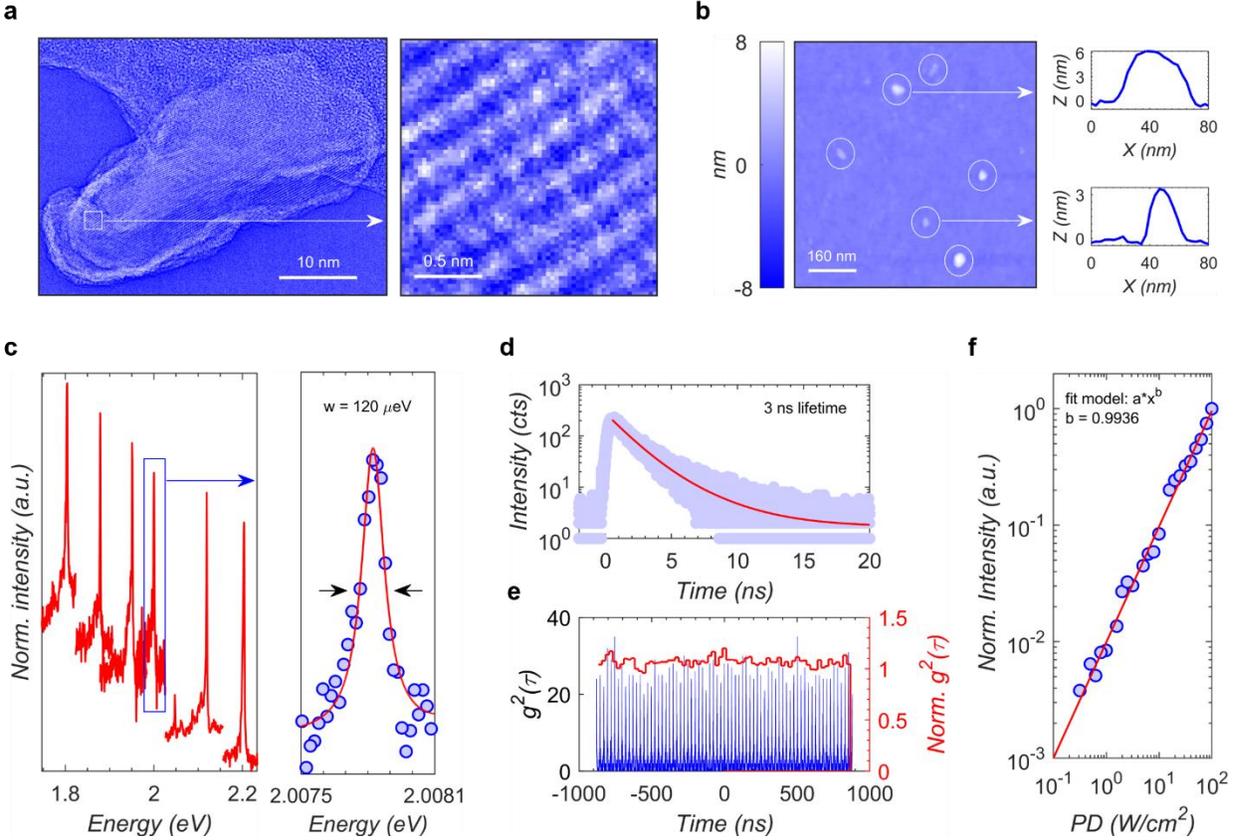

*Fig. 1: Ultra-sharp PL from nanoflakes. (a) HR-TEM image of a nanoflake with inset showing a zoom-in of lattice crystallinity. (b) Scanning probe image shows the size & thickness of deposited nanoflakes. (c) PL from six different nanoflakes showing a peak range from 1.8 – 2.2 eV, and a resolution limited linewidth of 120 μeV. (d) PL lifetime of 3 ns. (e) Second-order autocorrelation showing Poissonian photon-statistics. (f) A linear dependence to exitation power density.*

First-principles based calculations on bulk $NiPS_3$ show that exciton states lie in the energy range of 1.0 to 2.2 eV with their characters (*i.e.* spatial extent of the electron and hole wave functions) sensitively varying from *d-d* (*i.e.* electron and hole localized predominantly on Ni site) to charge-transfer (*i.e.* electrons and holes localized on different magnetic sublattices) depending on their energies.[18] Notably, a region of high energy states with charge-transfer characteristics is observed in 1.8 to 2.2 eV in-agreement with absorption experiments.[19] While earlier low temperature PL studies do not report any emission in this energy range, our PL data on bulk $NiPS_3$ (Fig. SI-4) reveals the emergence of a new emission peak at 2.1 eV when the laser excitation energy is increased to 3.06 eV. Since the emission energies of our NFs fall in this energy range, the sharp

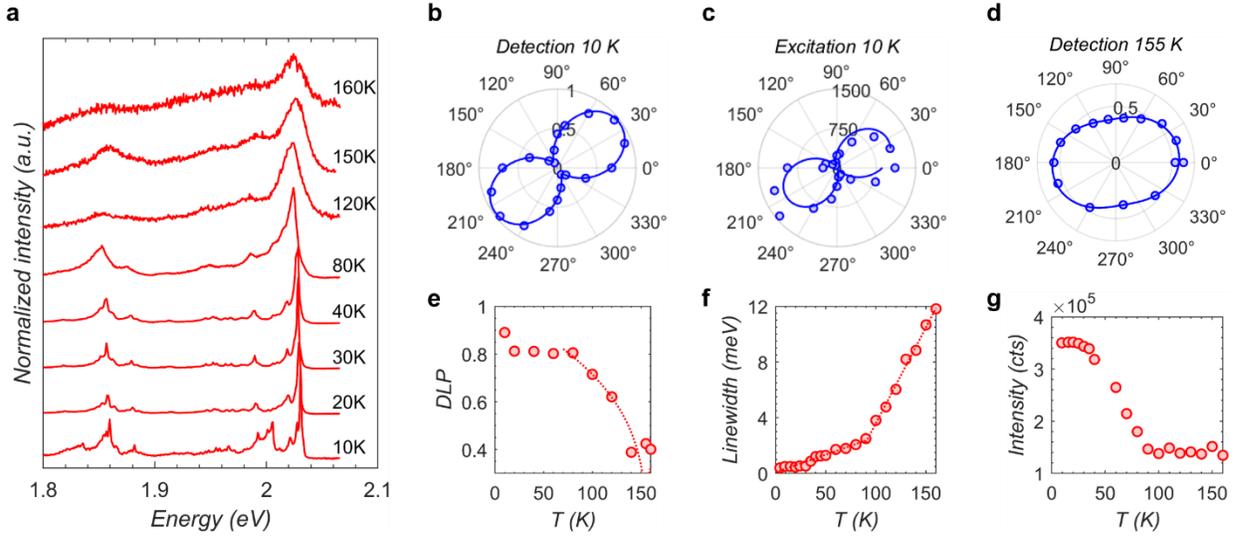

*Fig. 2: Spin-correlated PL in nanoflakes. (a) Temperature-dependent PL from a nanoflake (b-c) Polar plot showing a near unity degree of linear polarization at 10 K under detection and excitation polarization analysis. (d) Polar plot showing a drop in degree of linear polarization to 40% at 155 K near $T_N$. (e-g) Degree of linear polarization, linewidth and intensity for different temperatures.*

PL peaks of NFs are tentatively attributed to the activation of charge-transfer exciton states by edge effects that are known to dramatically alter electrical and optical properties of 2D materials upon reduction of their lateral dimensions (*e.g.* reconstruction of dangling-bonds at the edge). As the assignment of bulk SO-X is under debate with multiple possibilities reported such as Zhang-Rice singlet[21] or self-doping induced high density traps[22] or defect-bound states[26], our PL studies on bulk and NFs here show that excitons of different binding characters co-exist in $NiPS_3$ in agreement with our theoretical prediction.[18] Further, our following studies reveal that the PL emissions from NFs display clear signatures of intertwining charge, spin, and lattice degrees of freedom.

**Spin-correlated exciton**

Our temperature-dependent polarization resolved PL studies (Fig. 2a) reveal that the NFs display a near-unity linearly polarized PL under both emission and excitation polarization analysis (Fig. 2b,c), whereas SO-X only shows the former.[22] While the SO-X emission vanishes around 100 K, a very weak PL signal is still emitted near the Néel temperature giving us access to the degree of linear polarization (DLP) across the magnetic transition temperature. Strong linear polarization is observed at 10 K and is effectively constant (~90%-80%) between 20 K and 90 K. At 90 K, the DLP begins to precipitously decrease to 40% at ~150 K (Fig. 2d,e). The drop in linear polarization is fit to $|1 - \frac{T}{T_N}|^{2\beta}$ yielding a critical exponent β = 0.21±0.04, falling within the window of 0.1-to-0.25 expected for a 2D XY model.[22,27] The residual DLP above $T_N$ can be driven by short-range AFM order with the weak PL above $T_N$ similar to the persistence of the charge-transfer gap in bulk.[16,17] An analysis of the linewidth also reveals that while the linewidth increases slowly with a slope of 0.025 meV/K from 10-to-90 K, the slope drastically increases to 0.14 meV/K for 90-to-

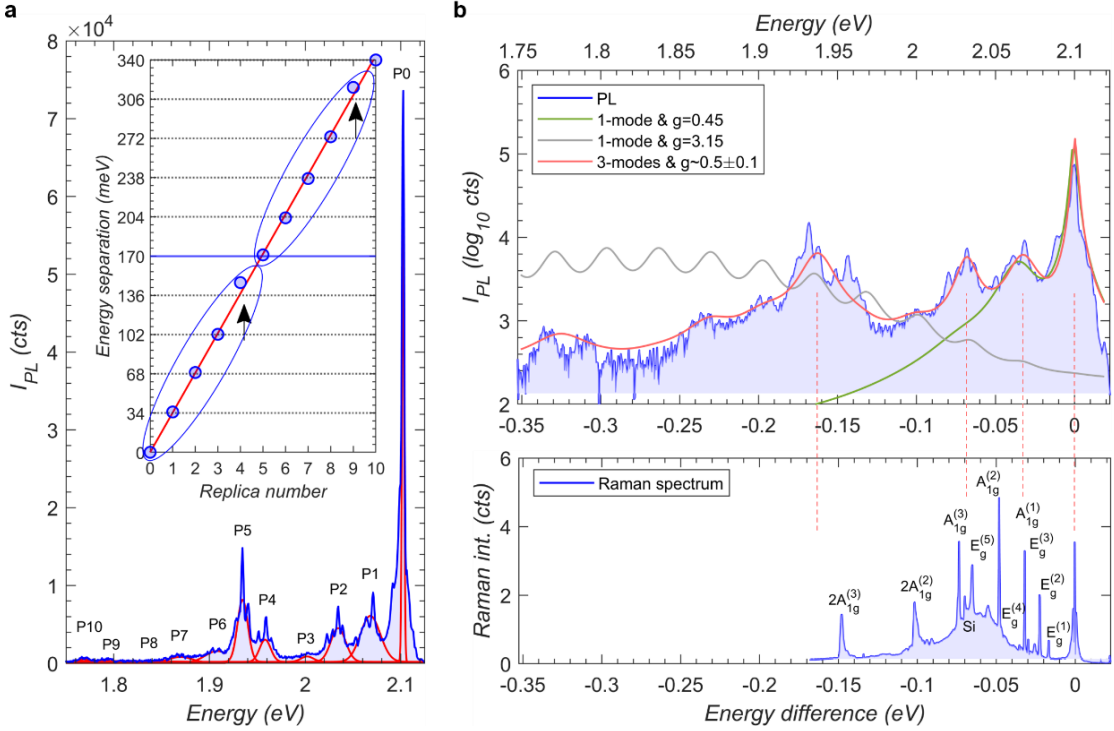

*Fig. 3: Exciton-phonon bound states in nanoflakes.* *(a) Low-resolution grating spectrum of a nanoflake showing the main peak P0 and 10 tail peaks labelled from P1 to P10. Inset: Tail peak positions from the central emission energy of P0 fitted with a linear function giving a slope of ~34 meV. Arrows show that P4 and P9 slightly deviate from the linear function. Ellipsoid shows the pattern of P0-P4 repeats at P5-P9. (b) PL lineshape (blue) fitted with the Holstein model accounting for 3 phonon modes with g~0.5±0.1 (red). For the lowest phonon mode, we calculated 1 phonon mode lineshapes using g=0.45 (green), and g=3.15 (gray). Raman spectrum of bulk NiPS$_3$ is also shown.*

160 K, reflecting the underlying AFM phase transition (Fig. 2f). The integrated intensity shows a similar temperature-dependence, with decreased intensities beyond 90 K (Fig. 2g). These behaviors are perfectly in line with those reported for SO-X,[22] and thus provide evidence that the ultra-sharp PL emission of our NF is strongly correlated to the spins of NiPS$_3$'s zig-zag antiferromagnetic order.

**Polaronic exciton coupled to multiple phonon modes**

The ultra-sharp PL emission peak (P0) from the NFs at cryogenic temperatures displays a low energy tail that is composed of up to 10 spectral peaks (P1-P10) spread over 340 meV (Fig. 3a collected at 10 K). See Supplementary Information S.4 for low energy tail peaks displaying similar polarization orientation as the main peak. Peak energies are linear in replica number with a slope of ~34 meV (Fig. 3a inset), and thus appear similar to the exciton-$A_{1g}^{(1)}$-phonon bound states reported in linear dichroism[20] and ultrafast spectroscopy[28] studies of the high energy range of 1.5 – 2 eV. Yet an adapted Holstein exciton-polaron model (see Supplementary Information S.5) for a single $A_{1g}^{(1)}$ phonon coupled to an exciton with a moderate coupling $g = 0.45$ (Fig. 3b, green line) can only reproduce the spectral lineshape up to P1 with the intensity monotonically decreasing to zero for larger energy differences. Increasing the coupling strength to $g = \sqrt{9.9} \approx 3.15$, as used in Ref[28], yields a Gaussian envelope of replicas that almost nullifies the zero-phonon line peak P0 in contrast to the experimental spectra (Fig. 3b, gray line). A closer inspection of the PL spectra for multiple NFs (Fig. SI-5,6) reveals a distinctive modulation in intensities of the replicas, where the set of peaks from P0 to P4 repeat at least three times over the full energy range. Since a single phonon model does not reasonably reproduce this peak pattern and the Raman spectrum of bulk NiPS$_3$ reveals 8 independent Raman active phonon modes (Fig. 3b and Fig. SI-7), we explore the possibility that the replicas are characterized by coupling to multiple phonon modes of NiPS$_3$. Here, we find a minimum of 3 phonon modes are required to fit the intensity modulation pattern of the phonon replicas. Specifically, phonon energies of 32.8, 68.5 and 163 meV with moderate coupling strength of g~0.5±0.1 generate the red line in Fig. 3b. While the energies of the first two modes are in good accord with the $A_{1g}^{(1)}$ and $E_g^{(5)}$ or $A_{1g}^{(3)}$ modes, respectively, the third mode does not match with any of the fundamental Raman modes of NiPS$_3$. We tentatively assign this phonon to a Raman inactive mode that becomes bright only after coupling to the exciton, similar to that observed in an another 2D magnet CrI$_3$.[29] Our analysis suggests a rich manifold of phonon modes coupled to the exciton in NiPS$_3$ resulting in the observed multiple phonon replicas, rather than a single strongly coupled phonon mode.

**Discrete spectral jumps and capacitor charging model**

The PL spectrum shown in Fig. 3a reveals that each peak is a superposition of sharp lines. We further investigate this by acquiring a series of 400 high-resolution PL spectra at 1s intervals (Fig. 4a). This shows the P0 sharp line undergoing discrete spectral jumps and those near P1 follow the magnitude and direction of the former. A time integrated PL spectrum is shown in Fig. 4b that displays a series of sharp PL peaks covering the entire spectral range of P0 to P5 with a comb-like pattern. The zoom-in view of this pattern (Fig. 4b inset) and histogram of consecutive peak-to-peak energy separation (Fig. 4c) further reveal that two adjacent peaks are separated by a regular energy step of $\delta_1 \sim 1.9 \pm 0.1$ meV with some peaks having a lower intensity shoulder separated by $\delta_2 \sim 0.75 \pm 0.1$ meV. These findings provide clear evidence that the spectral lines jump randomly in time but in regular steps of energy. Spectral jumps in other nanoscale light emitters such as quantum dots or single molecules are attributed to the Stark shift of excitonic emission resulting from the electric field of photo-excited carriers trapped in the surrounding.[30,31] Since the configuration and number of trapped carriers are random in both time and space, so too are the spectral jumps in most nanoscale emitters. Our observation of spectral jumps with regular energy intervals across different NFs (Fig. SI-6) suggests that electric field fluctuations responsible for the Stark shift of the emission lines are occurring in regular increments. To explain this, we propose a charging model (see Supplementary Information S.6) in which the electric field inside the NF is a consequence of excess charge carriers that are uniformly delocalized within the layers (Fig. 4d). As such, each individual layer of the NF contributes to a uniform electric field. Here, the charging event is attributed to the interlayer tunneling of energetic electrons or holes after photo-excitation while the carriers are still hot. This transfer of an electron to an adjacent layer while leaving the hole behind (or vice versa) gives the uniform field of a parallel plate capacitor straddling the barrier between layers which is on the order of $10^5$ V/m and sufficient enough to induce regular Stark shifts in the order of ~1 meV. To validate this model, we perform Kinetic Monte Carlo simulations. The results show reasonable agreement with the experimental data (Fig. SI-10,11,12) providing support for our attribution of discrete Stark shifts in energy steps, occurring randomly in time, to interlayer charge hopping dynamics creating a homogeneous electric field.

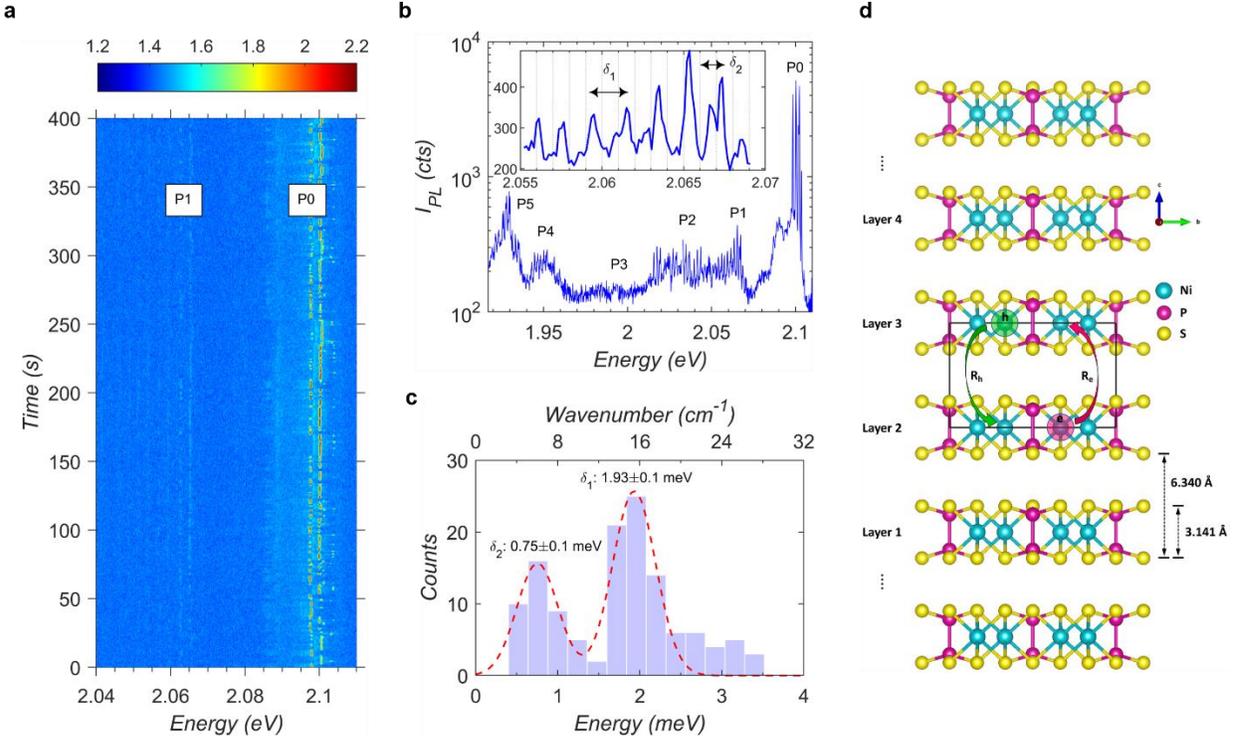

*Fig. 4: Spectral lines jump in regular steps of energies in nanoflakes. (a) A time dependent spectral image sequence of PL collected from the same nanoflake shown in Fig. 3 with 1s each acquisition under a high-resolution grating spectrometer. P0 sharp line undergoes discrete spectral jumps which are followed in magnitude and direction by the P1 sharp lines. (b) A time integrated spectral plot collected with high-resolution grating spectrometer displayed in log scale. Discrete jumps are visible over the entire span of P0-to-P5 and the inset shows a small window with noticeable equidistant consecutive peak separations marked as δ1 and δ2. (c) Histogram of consecutive peak difference gives a value of ~1.9 meV for δ1 and ~0.7 meV for δ2. (d) A schematic of our parallel-capacitor charging model where inter-layer electron (hole) hopping with the rate $R_e$ ($R_h$) upon photo-excitation creates a homogeneous electric field.*

**Discussion**

In summary, our findings here reveal that the spin-correlated excitonic signatures seen in bulk are preserved in NFs comprising only tens of nanometer lateral sizes prepared by chemical synthesis. The excitonic signatures elaborated in the Results show that the excitons in nanoscale correlated material are quite different than the excitons in conventional nanoscale semiconductors exhibiting weak correlations. While we interpret the distinct PL lineshape of our NFs using the Holstein exciton-polaron model with evidence for exciton coupled to multiple phonon modes with only moderate coupling strengths each, which differs from the bulk exciton coupled to one phonon mode with the strongest coupling strength reported in any 2D semiconductors,[28] the effect of phonon-driven optical transduction conveyed in the bulk exciton would also be applicable here. Though the spectral diffusion of exciton observed universally across all nanoscale light emitters is considered an avoidable optical behavior, the fundamental knowledge of the physical mechanism behind the exciton formation got better understood, for instance the excited state process in quantum dots with the efforts for suppressing the blinking[32] resulted in the development of nanoscale photonic devices. Similarly, our report of the unusual comb-like Stark shift interpreted as the discrete energetic comb arising from the discretization of charges stored in each layers producing uniform field with the aid of a capacitor charging model shows that the charging scenario is exclusive to the nanoscale correlated antiferromagnetic material with possibilities for external field control. Our PL results on bulk and NFs show that one material of $NiPS_3$ exhibits different co-existing excitonic binding characters such as charge-transfer, *d-d*, and spin-orbit entanglement. The selective activation of one of the above characters, charge-transfer like inter-magnetic sublattice exciton, here in NFs shows that lateral confinement could be considered as a tool to control excitonic behavior. The interplay of different degrees of freedom influencing the excitonic signatures also indicate the tuning of optical properties is possible by gradient chemical composition and doping. While this unlocking of high-energy exciton in the NFs could be compared to the similar activation of different PL peak energies in surface modified nanodiamond counterpart of bulk diamond[33] or nanoribbon counterpart of graphene[34], here the combination of smaller lateral dimension, correlated antiferromagnetic structure and van der Waals interlayer stacking configuration contribute to the uniqueness in the observed excitonic signatures.

**Methods:**

**Sample Preparation:**

Our solution phase synthesis relies on the metathesis reaction of highly ionic $X_4P_2S_6$ (x= Na, Li) with common nickel salts ($NiCl_2$, $Ni(NO_3)_2$), and the replacement of the oxygen around the nickel by octahedral $P_2S_6^{4-}$ anion. This approach results in $NiPS_3$ crystalline domains of 3 nm in diameter in an amorphous matrix. This is followed by a high temperature annealing at 350 °C for 30 minutes resulting in nanoflakes of $NiPS_3$ which are then dissolved into a colloidal form with deionized water. A 100 µL of the suspended nanoflake solution is dropcasted onto $Si/SiO_2$ substrates for PL measurements on individual nanoflakes.

**PL Measurements:**

A) Micro-photoluminescence setup

A continuous wave solid state laser with an energy of 2.33 eV is used for sample excitation. The laser is reflected through a dichroic beamsplitter and then focused onto the sample (placed in Oxford MicrostatHires liquid Helium flow cryostat) to a diffraction limited spot size using a 50×, 0.7 NA Olympus objective microscope, which is used to both excite the sample and collect the PL. Collected photons after passing through the same dichroic beamsplitter and a 545 nm long-pass filter go to a spectrometer + LN-cooled CCD combo (Acton SP2300i/Pylon 100BR or Acton SP2500i/Pylon 2KB) with 300 g/mm or 1200 g/mm or 2400 g/mm gratings.

B) Polarization-resolved spectroscopy

For polarization-resolved spectroscopy, a non-polarizing beamsplitter is used in place of dichroic beamsplitter and a Wollaston prism is placed just before the spectrometer + CCD combo to split the signal into *s* and *p* polarized components. To investigate the PL polarization, a half-wave plate is used as an analyzer either in the detection path after the beamsplitter or the excitation path before the beamsplitter.

C) Time-resolved spectroscopy

For time-resolved photoluminescence (TRPL) lifetime and second order autocorrelation experiments, a picosecond pulsed laser centered at 2.33 eV with a tunable repetition rate from 1-40 MhZ is used for sample excitation and a Hanbury Brown-Twiss setup consisting of a 50:50 beamsplitter & two single photon avalanche photodiodes (Excelitas SPCM-AQRH-14) with the timing resolution of 300 ps are used. TRPL is analyzed using a TCSPC module (Picoquant Hydraharp 400).

**HR-TEM Measurements:**

High resolution transmission electron microscopy (HR-TEM) is performed on a monochromated and aberration corrected FEI Titan operating at 300 keV.

**Scanning Probe Measurements:**

The surface morphology of individual nanoflakes is measured using an atomic force microscope (Bruker Dimension Icon) operating in tapping mode under ambient conditions at room temperature.

**Raman Measurements:**

Raman spectroscopic measurements are conducted in backscattering/reflection mode using a 532.3 nm excitation (Oxxius LCX-532S-100, 100 mW continuous wave single longitudinal mode diode pumped solid state laser) in a Horiba LabRAM HR Evolution microscope fitted with volume Bragg gratings. Experiments are conducted in an Oxford Microstat liquid helium cryostat using a 60x, 0.7 NA glass corrected semi-apochromat objective (Olympus LUCPLFLN60X) and the instrument is configured using an 1800 l/mm grating and a 50 µm confocal hole.

**Theoretical Calculations:**

Analytical solution of Holstein model for exciton-polaron state accounting for the exciton coupling with three phonon modes (see Supplementary Information S.5) are used to extract exciton-phonon coupling strength and phonon frequencies. A parallel capacitor model is proposed to capture the NF charging effects (see Supplementary Information S.6) and this is combined with the Kinetic Monte Carlo simulations and the Holstein exciton-polaron model to explain the PL spectral diffusion and spectral fine structure, respectively.

**Data Availability:**

The data that support the findings of this study are available within the main text and Supplementary Information. Any other relevant data are available from the corresponding authors upon reasonable request.

**References:**

1    Koch, S. W., Kira, M., Khitrova, G. & Gibbs, H. M. Semiconductor excitons in new light. *Nat. Mater.* **5**, 523-531 (2006).
2    Basov, D. N., Averitt, R. D. & Hsieh, D. Towards properties on demand in quantum materials. *Nat. Mater.* **16**, 1077-1088 (2017).
3    Giustino, F. *et al.* The 2021 quantum materials roadmap. *J. Phys. Mater.* **3**, 042006 (2020).
4    Dagotto, E. Complexity in strongly correlated electronic systems. *Science* **309**, 257-262 (2005).
5    Ma, L. *et al.* Strongly correlated excitonic insulator in atomic double layers. *Nature* **598**, 585-589 (2021).
6    Morita, Y., Yoshioka, K. & Kuwata-Gonokami, M. Observation of Bose-Einstein condensates of excitons in a bulk semiconductor. *Nat. Commun.* **13**, 5388 (2022).
7    Onga, M., Zhang, Y., Ideue, T. & Iwasa, Y. Exciton Hall effect in monolayer $MoS_2$. *Nat. Mater.* **16**, 1193-1197 (2017).
8    Seradjeh, B., Moore, J. E. & Franz, M. Exciton condensation and charge fractionalization in a topological insulator film. *Phys. Rev. Lett.* **103**, 066402 (2009).
9    Kavokin, A. & Lagoudakis, P. Exciton-polariton condensates: Exciton-mediated superconductivity. *Nat. Mater.* **15**, 599-600 (2016).
10   Kogar, A. *et al.* Signatures of exciton condensation in a transition metal dichalcogenide. *Science* **358**, 1314-1317 (2017).
11   Mori, R. *et al.* Spin-polarized spatially indirect excitons in a topological insulator. *Nature* **614**, 249-255 (2023).
12   Kim, D. *et al.* Robust Interlayer-Coherent Quantum Hall States in Twisted Bilayer Graphene. *Nano Lett.* **23**, 163-169 (2023).
13   Kurebayashi, H., Garcia, J. H., Khan, S., Sinova, J. & Roche, S. Magnetism, symmetry and spin transport in van der Waals layered systems. *Nat. Rev. Phys.* **4**, 150-166 (2022).
14   Wang, Q. H. *et al.* The Magnetic Genome of Two-Dimensional van der Waals Materials. *ACS Nano* **16**, 6960-7079 (2022).
15   Burch, K. S., Mandrus, D. & Park, J. G. Magnetism in two-dimensional van der Waals materials. *Nature* **563**, 47-52 (2018).
16   Kim, S. Y. *et al.* Charge-Spin Correlation in van der Waals Antiferromagnet $NiPS_3$. *Phys. Rev. Lett.* **120**, 136402 (2018).

17  Lane, C. & Zhu, J.-X. Thickness dependence of electronic structure and optical properties of a correlated van der Waals antiferromagnetic NiPS3 thin film. *Phys. Rev. B* **102**, 075124 (2020).
18  Lane, C. & Zhu, J.-X. An ab initio study of electron-hole pairs in a correlated van der Waals antiferromagnet: NiPS3. Preprint at https://doi.org:10.48550/arXiv.2209.13051 (2022).
19  Banda, E. J. K. B. Optical absorption of NiPS3 in the near-infrared, visible and near-ultraviolet regions. *J. Phys. C: Solid State Phys.* **19**, 7329-7335 (1986).
20  Hwangbo, K. *et al.* Highly anisotropic excitons and multiple phonon bound states in a van der Waals antiferromagnetic insulator. *Nat. Nanotechnol.* **16**, 655-660 (2021).
21  Kang, S. *et al.* Coherent many-body exciton in van der Waals antiferromagnet NiPS3. *Nature* **583**, 785-789 (2020).
22  Wang, X. *et al.* Spin-induced linear polarization of photoluminescence in antiferromagnetic van der Waals crystals. *Nat. Mater.* **20**, 964-970 (2021).
23  Dirnberger, F. *et al.* Spin-correlated exciton-polaritons in a van der Waals magnet. *Nat. Nanotechnol.* **17**, 1060-1064 (2022).
24  Belvin, C. A. *et al.* Exciton-driven antiferromagnetic metal in a correlated van der Waals insulator. *Nat. Commun.* **12**, 4837 (2021).
25  Scholes, G. D. & Rumbles, G. Excitons in nanoscale systems. *Nat. Mater.* **5**, 683-696 (2006).
26  Kim, D. S. *et al.* Anisotropic Excitons Reveal Local Spin Chain Directions in a van der Waals Antiferromagnet. *Adv. Mater.* **n/a**, 2206585 (2023).
27  Wildes, A. R. *et al.* Magnetic structure of the quasi-two-dimensional antiferromagnet NiPS3. *Phys. Rev. B* **92**, 224408 (2015).
28  Ergecen, E. *et al.* Magnetically brightened dark electron-phonon bound states in a van der Waals antiferromagnet. *Nat. Commun.* **13**, 98 (2022).
29  Jin, W. *et al.* Observation of the polaronic character of excitons in a two-dimensional semiconducting magnet CrI3. *Nat. Commun.* **11**, 4780 (2020).
30  Frantsuzov, P., Kuno, M., Jankó, B. & Marcus, R. A. Universal emission intermittency in quantum dots, nanorods and nanowires. *Nat. Phys.* **4**, 519-522 (2008).
31  Empedocles, S. A. & Bawendi, M. G. Influence of Spectral Diffusion on the Line Shapes of Single CdSe Nanocrystallite Quantum Dots. *J. Phys. Chem. B* **103**, 1826-1830 (1999).
32  Galland, C. *et al.* Two types of luminescence blinking revealed by spectroelectrochemistry of single quantum dots. *Nature* **479**, 203-207 (2011).
33  Mochalin, V. N., Shenderova, O., Ho, D. & Gogotsi, Y. The properties and applications of nanodiamonds. *Nat. Nanotechnol.* **7**, 11-23 (2011).
34  Wang, H. *et al.* Graphene nanoribbons for quantum electronics. *Nat. Rev. Phys.* **3**, 791-802 (2021).

**Acknowledgements**

89233218CNA000001. H.H., V.C., C.R.D., D.P., C.A.L., J.-X.Z., M.A.C., S.A.I., and A.P. acknowledge primary support for the works from Laboratory Directed Research and Development (LDRD) program 20200104DR. S.A.I., V.C., and C.T.T. are also supported by LDRD 20220757ER. H.H., V.C., X.L., and H.Z. acknowledge partial support by Quantum Science Center, a National QIS Research Center supported by DOE, OS. M.T.P. acknowledges support from LDRD awards 20210782ER and 20210640ECR. A.C.J. is supported by DOE BES QIS program LANLE3QR.

**Author information**

Authors and Affiliations

**Center for Integrated Nanotechnologies, Materials Physics and Applications Division, Los Alamos National Laboratory, Los Alamos, New Mexico 87545, United States**
Vigneshwaran Chandrasekaran, Christopher R. DeLaney, David Parobek, Jian-Xin Zhu, Xiangzhi Li, Huan Zhao, Cong Tai Trinh, Marshall A. Campbell, Andrew C. Jones, John Watt, Michael T. Pettes, Sergei A. Ivanov, Han Htoon

**Theoretical Division, Los Alamos National Laboratory, Los Alamos, New Mexico 87545, United States**
Christopher A. Lane, Jian-Xin Zhu, Andrei Piryatinski

**Materials Science in Radiation and Dynamics Extremes, Materials Science and Technology Division, Los Alamos National Laboratory, Los Alamos, New Mexico 87545, United States**
Matthew M. Schneider

**Department of Physics and Astronomy, University of New Mexico, Albuquerque, New Mexico 87131, United States**
David H. Dunlap

Contributions
H.H. conceived and led the experiment. V.C. performed the optical experiments, analyzed the data, plotted the graphics, and drafted the manuscript. C.R.D., D.P., and S.A.I. synthesized the nanoflakes. M.M.S., and J.W. obtained the TEM images. M.A.C., V.C., and M.T.P. performed the Raman spectroscopy. X.L., H.Z., and C.T.T. assisted V.C. in the PL experiments. A.C.J. assisted V.C. in the scanning probe measurements. C.A.L., and J.-X.Z. provided inputs for theoretical models. A.P., and D.H.D. provided theoretical grounds for interpreting the spectroscopic data using the Holstein exciton-polaron model along with the charge fluctuating capacitor model. V.C., and H.H. finalized the manuscript with inputs from C.A.L., S.A.I., A.P., and D.H.D.

Corresponding authors
Correspondence to vcha@lanl.gov or dunlap@unm.edu or htoon@lanl.gov

# Supplementary Information for

# Correlated Excitonic Signatures in a Nanoscale van der Waals Antiferromagnet

Vigneshwaran Chandrasekaran,[1*] Christopher R. DeLaney,[1] David Parobek,[1] Christopher A. Lane,[2] Jian-Xin Zhu,[1,2] Xiangzhi Li,[1] Huan Zhao,[1] Cong Tai Trinh,[1] Marshall A. Campbell,[1] Andrew C. Jones,[1] Matthew M. Schneider,[3] John Watt,[1] Michael T. Pettes,[1] Sergei A. Ivanov,[1] Andrei Piryatinski,[2] David H. Dunlap,[4*] Han Htoon[1*]

[1] Center for Integrated Nanotechnologies, Materials Physics and Applications Division, Los Alamos National Laboratory, Los Alamos, New Mexico 87545, United States

[2] Theoretical Division, Los Alamos National Laboratory, Los Alamos, New Mexico 87545, United States

[3] Materials Science in Radiation and Dynamics Extremes, Materials Science and Technology Division, Los Alamos National Laboratory, Los Alamos, New Mexico 87545, United States

[4] Department of Physics and Astronomy, University of New Mexico, Albuquerque, New Mexico 87131, United States

## Contents

## S.1. Characterization of Chemically Synthesized NiPS₃ Nanoflakes

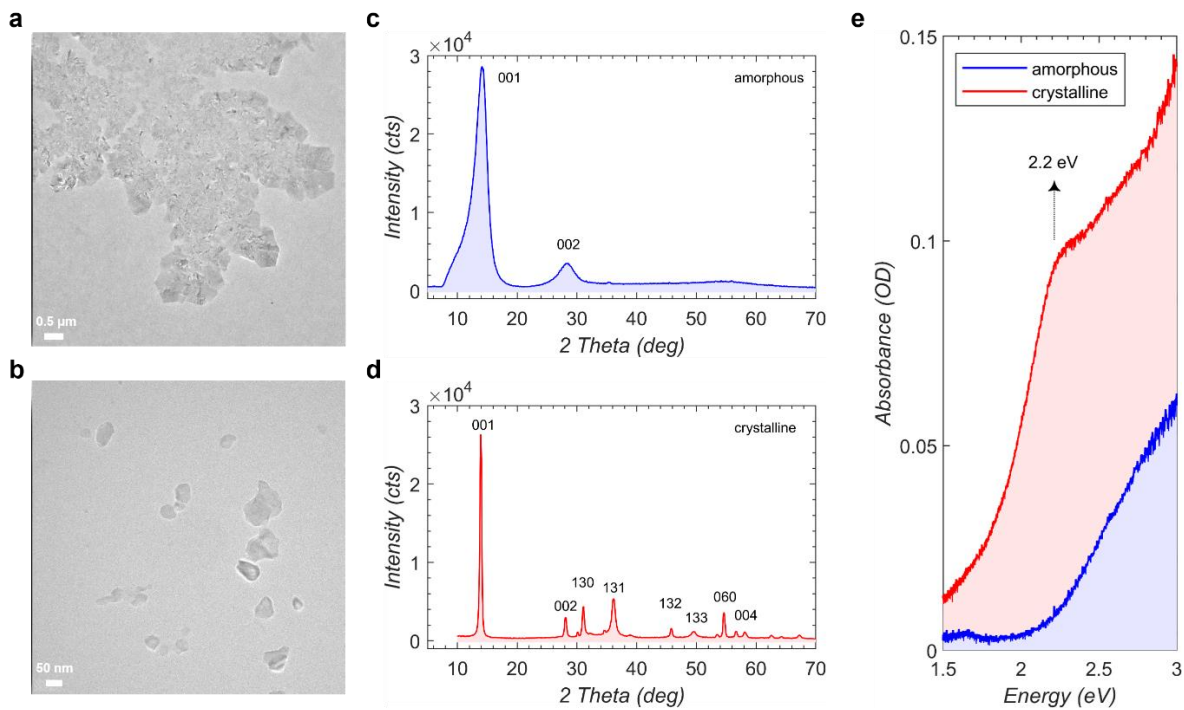

**Fig. SI-1: TEM, XRD and absorption spectra of amorphous and crystalline nanoflakes.** (a,b) TEM images of nanoflakes in amorphous and crystalline phases respectively. Size range of crystalline nanoflakes is 20-100 nm. (c,d) XRD obtained from amorphous and crystalline nanoflakes of NiPS₃. (e) Absorption spectra of amorphous and crystalline nanoflakes of NiPS₃. Crystalline sample shows a pronounced peak around 2.2 eV.

## S.2. Spin-orbit Entangled Exciton Emission in Micrometer-sized Flakes

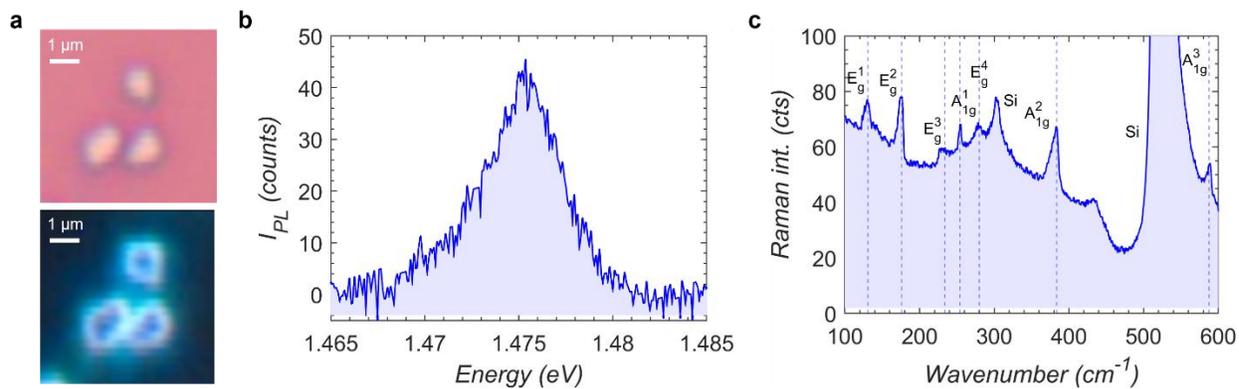

**Fig. SI-2: Spin-orbit entangled exciton emission in micrometer-sized flakes.** (a) Bright field and dark field images of ~1 μm lateral size bulk flakes of $NiPS_3$ prepared by micro-mechanical exfoliation technique. (b) PL of the flake showing the spin-orbit entangled exciton emission at ~1.475 eV. (c) Raman signal of respective flake.

## S.3. Excitation Power and Laser Energy Dependence of PL

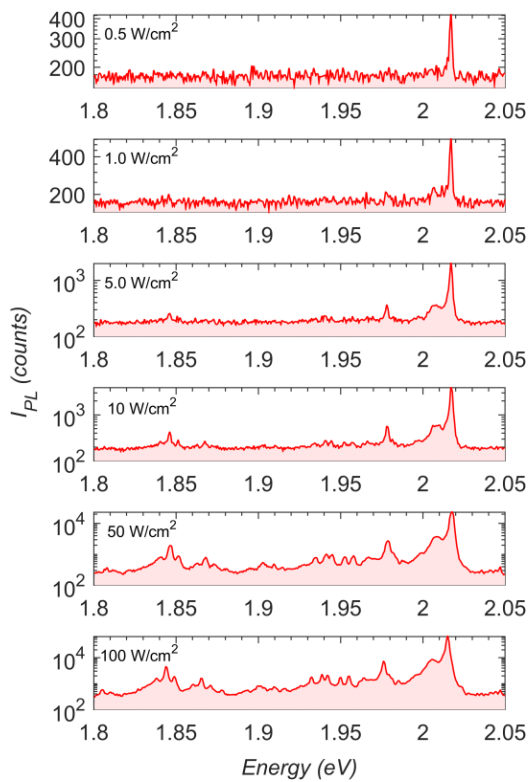

**Fig. SI-3: Excitation power dependence of PL in a nanoflake.** PL from a nanoflake under different excitation power density. Low energy tail peaks become visible at higher excitation power density.

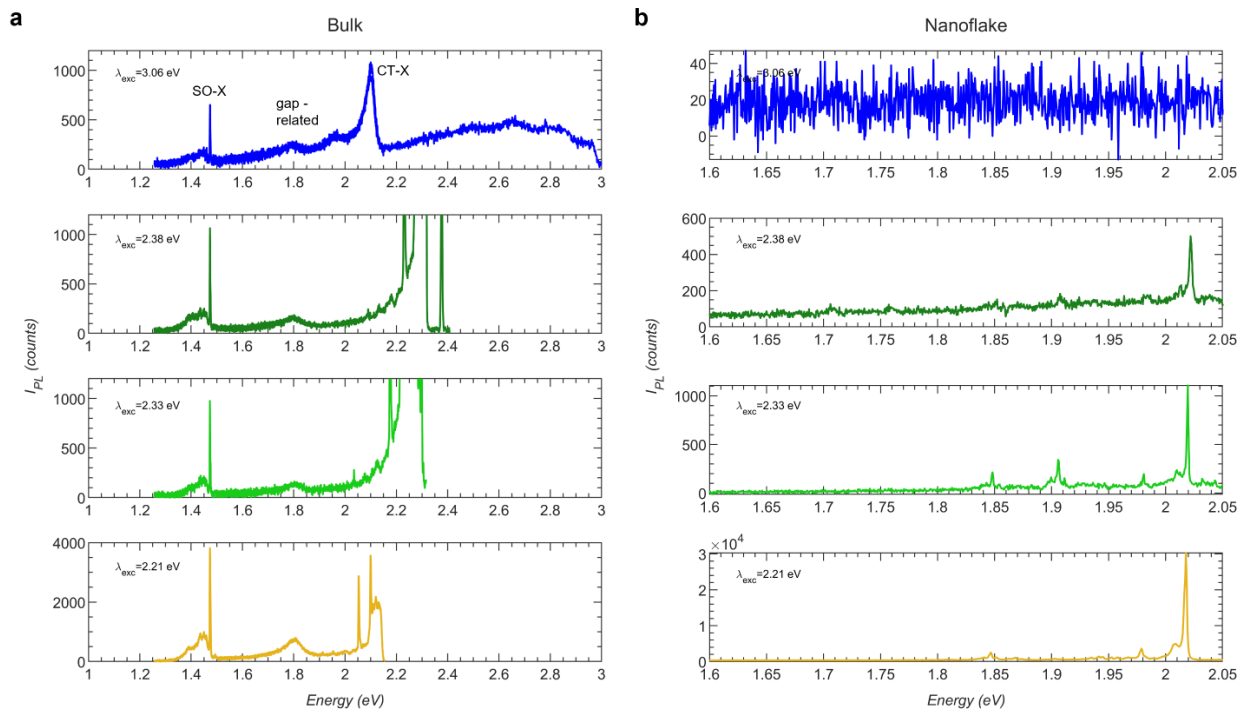

**Fig. SI-4: Excitation laser energy dependence of PL in bulk & a nanoflake.** PL collected with different excitation laser energies under confocal excitation with a diffraction limited spot size for (a) bulk and (b) a nanoflake $NiPS_3$ with the same power density of about 70 W/cm$^2$. Higher energy excitation at 3.06 eV for bulk has the SO-X exciton at 1.47 eV whose peak PL intensity is weaker than another peak at 2.1 eV. Since this peak energy falls in the charge-transfer peak region reported for $NiPS_3$, we assign it to a charge-transfer exciton (CT-X). CT-X has a side peak at 1.95 eV. Another peak at 1.8 eV is also observed which we assign to an exciton related to bandgap of $NiPS_3$. Under lower energy excitations between 2.38 eV and 2.21 eV, the CT-X peak at 2.1 eV is vanished leaving the SO-X at 1.47 eV and gap-related peak at 1.8 eV. Nanoflakes do not have SO-X exciton but only has the peak in visible region that becomes stronger for near-resonant excitation closer to charge-transfer state at 2.2 eV.

## S.4. Exciton-phonon Bound States

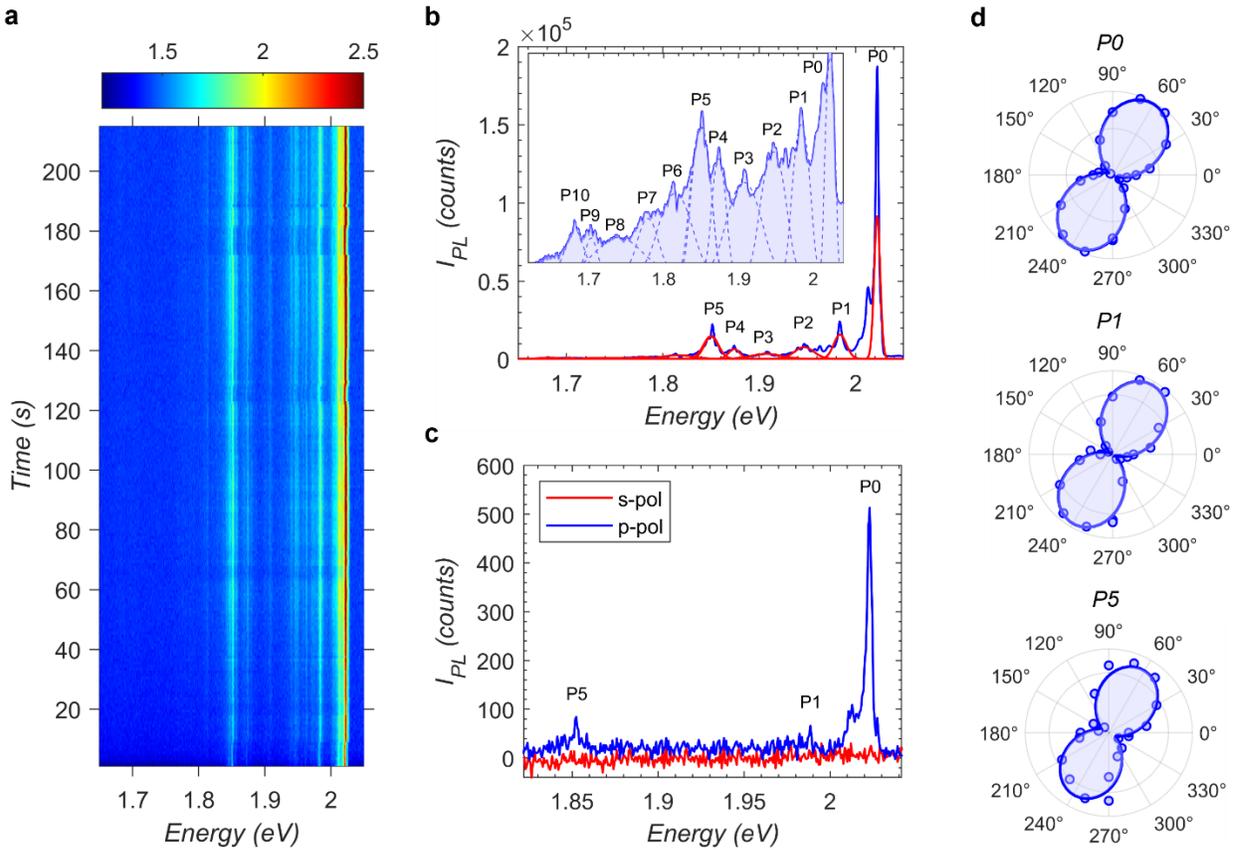

**Fig. SI-5: Low energy tail PL peaks follow the main PL peak in a nanoflake.** (a) A time dependent spectral image sequence of PL collected from a nanoflake with 1s each acquisition. This shows the low energy tail peaks following the direction of main peak. (b) Integrated spectrum of the same nanoflake showing the main peak P0 and 10 low energy tail peaks labelled from P1 to P10. Inset shown in log plot. (c) Polarization resolved spectra of the same nanoflake with 3 peaks P0, P1 and P5 visible. (d) Polarplot of the respective peaks displaying a strong linear polarization oriented in the same direction.

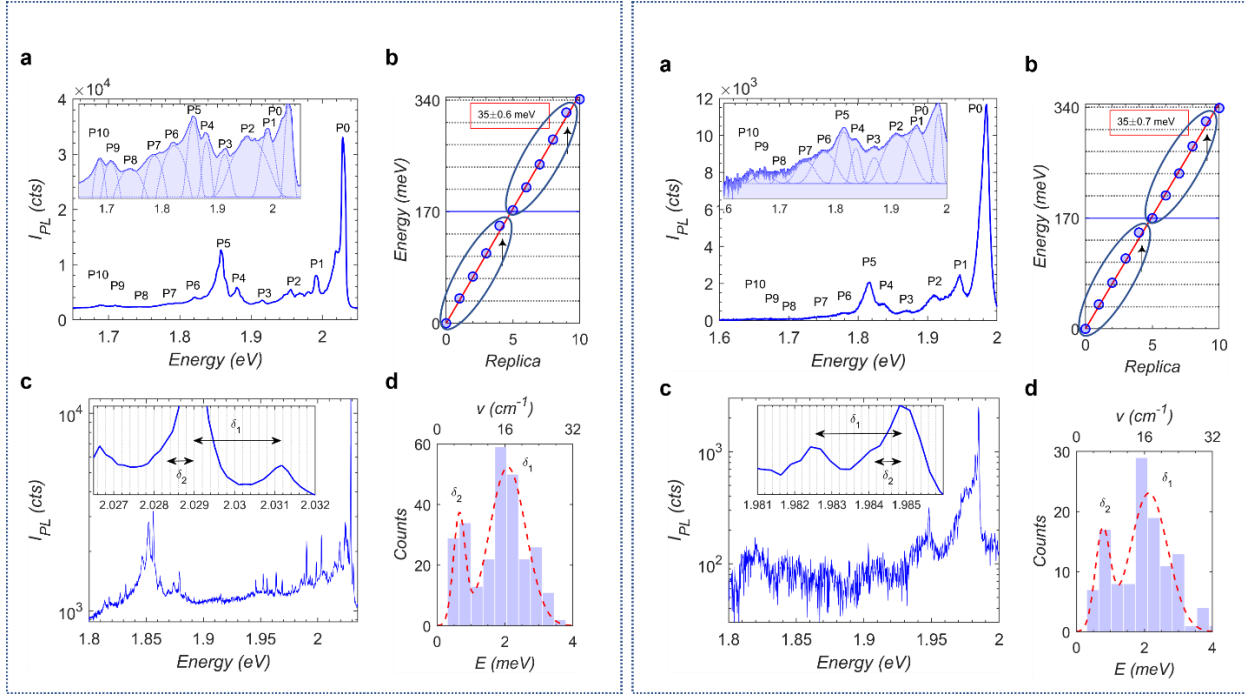

**Fig. SI-6: Low- and high-resolution spectra of 2 different nanoflakes.** (a) Low-resolution spectrum of a nanoflake showing the main peak P0 and 10 low energy tail peaks labelled from P1 to P10. Inset shown in log plot. (b) Tail peak positions from the central emission energy of P0 fitted with a linear function giving a slope of ~35 meV. Arrows show that P4 and P9 slightly deviate from the linear function. Ellipsoid shows the pattern of P0-P4 repeats at P5-P9. (c) High-resolution spectrum of the same nanoflake. Inset shows a zoomed-in portion with noticeable equidistant consecutive peak separations marked as δ1 and δ2. (d) Histogram of consecutive peak difference gives a value of ~2 meV for δ1 and ~0.7 meV for δ2.

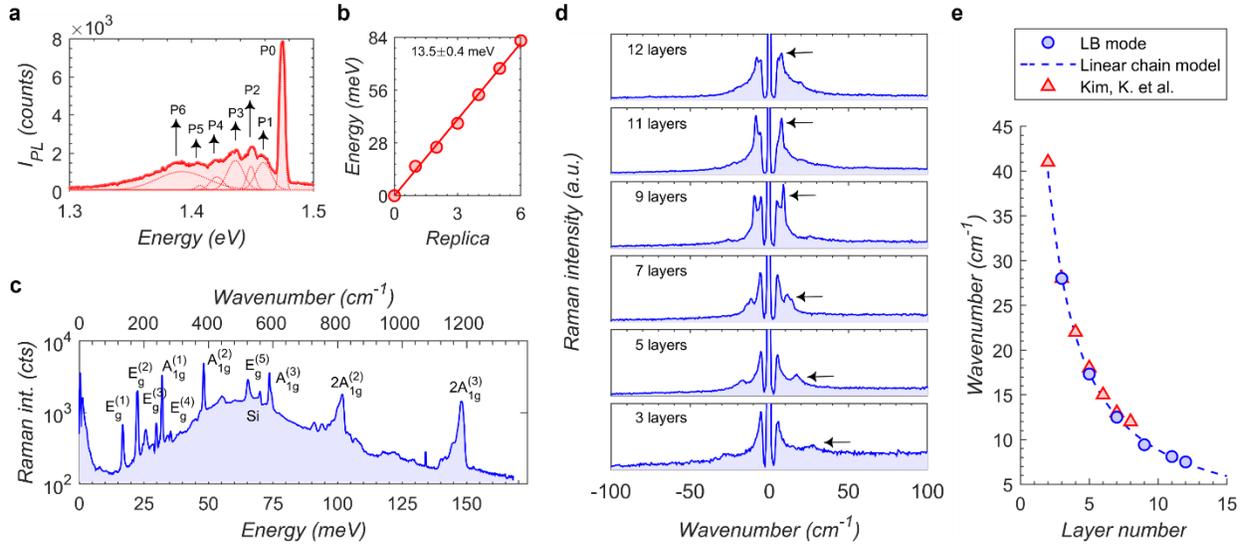

**Fig. SI-7: Exciton-phonon bound states & Raman phonon modes in bulk NiPS$_3$.** (a) PL spectrum from a bulk NiPS$_3$ showing the spin-orbit entangled exciton peak P0 at 1.475 eV and 6 low energy tail peaks marked as P1 to P6. (b) Tail peak positions from the central emission energy of P0 fitted with a linear function giving a slope of ~13.5 meV. (c) High frequency Raman phonon modes from a 12-layer thick NiPS$_3$. (d) Low frequency Raman phonon mode for specified layer thicknesses. Arrows point the position of LB mode. (e) LB mode dependence to layer number is fitted with a linear chain model. All data are collected at 4.2 K. Data of Kim, K. et al. obtained from *Nat Commun* **10**, 345 (2019).

## S.5. Holstein Model for Exciton-Polaron Quasiparticle and Associated PL Lineshape

Let us introduce the Holstein model assuming on site interaction of the exciton and phonon states described by Bose {creation, annihilation} operators $\{\hat{a}_n, \hat{a}_n^\dagger\}$ and $\{\hat{b}_q, \hat{b}_q^\dagger\}$, respectively. Notice that the exciton operators are presented in the real space whereas the phonon operators are in momentum space $q$. The model Hamiltonian in such a representation reads

$$\hat{H} = \hbar \sum_n [\Delta_0 + \sum_q g_q \omega_q e^{iqn}(\hat{b}_q + \hat{b}_{-q}^\dagger)] \hat{a}_n^\dagger \hat{a}_n + \hbar \sum_q \omega_q \hat{b}_q^\dagger \hat{b}_q , \qquad (1.1)$$

with index $n$ running over the exciton delocalization range, $\hbar\Delta_0$ denoting bare exciton transition energy, $\hbar\omega_q$ standing for the phonon energy, and $g_q$ being dimensionless exciton-phonon coupling parameter represented in the momentum space. Following standard approach, we perform the Lang-Firsov (displacement) transformation

$$\hat{D} = exp\{\sum_{nq} g_q e^{iqn}(\hat{b}_q - \hat{b}_{-q}^\dagger) \hat{a}_n^\dagger \hat{a}_n\} \qquad (1.2)$$

of the Hamiltonian (1.1), i.e., $\hat{H}' = \hat{D}\hat{H}\hat{D}^{-1}$. The transformation eliminates the exciton-phonon coupling term resulting in the Hamiltonian for the uncoupled dressed exciton and phonon states

$$\hat{H}' = \hbar \sum_n (\Delta_0 - \sum_{q,n'} g_q^2 \omega_q e^{iqn'} \hat{A}_{n'}^\dagger \hat{A}_{n'}) \hat{A}_n^\dagger \hat{A}_n + \hbar \sum_q \omega_q \hat{B}_q^\dagger \hat{B}_q. \qquad (1.3)$$

Here, the transformed exciton and phonon operators are

$$\hat{A}_n = \hat{D}\hat{a}_n \hat{D}^{-1} = \hat{a}_n e^{-\sum_q g_q e^{iqn}(\hat{b}_q - \hat{b}_{-q}^\dagger)} , \qquad (1.4)$$

$$\hat{B}_q = \hat{D}\hat{b}_q \hat{D}^{-1} = \hat{b}_q + \sum_n g_q e^{iqn} \hat{a}_n^\dagger \hat{a}_n, \qquad (1.5)$$

respectively. The inverse transformation gives the following relationship between bare exciton and phonon operators in term of the new operators

$$\hat{a}_n = \hat{A}_n e^{\sum_q g_q e^{iqn}(\hat{B}_q - \hat{B}_{-q}^\dagger)} , \qquad (1.6)$$

$$\hat{b}_q = \hat{B}_q - \sum_n g_q e^{iqn} \hat{A}_n^\dagger \hat{A}_n . \qquad (1.7)$$

To simplify the quartic term in the transformed Hamiltonian (1.3), we apply the mean-field approximation by introducing the on-site exciton density $\langle \hat{A}_n^\dagger \hat{A}_n \rangle$. This approximation results in the diagonalized Hamiltonian

$$\hat{H}_{MF} = \hbar\Delta \sum_n \hat{A}_n^\dagger \hat{A}_n + \hbar \sum_q \omega_q \hat{B}_q^\dagger \hat{B}_q, \qquad (1.8)$$

where the exciton polaron energy is $\Delta = (\Delta_0 - 2\sum_{qk} g_q^2 \omega_q \langle \hat{A}_k^\dagger \hat{A}_{k+q} \rangle)$ including the polaron shift $-2\sum_{qk} g_q^2 \omega_q \langle \hat{A}_k^\dagger \hat{A}_{k+q} \rangle$ proportional to the zero-time exciton correlation function, $\langle \hat{A}_k^\dagger \hat{A}_{k+q} \rangle$, in the momentum space. With the help of Eq. (1.8), we find the following representation of the bare exciton operator Eq. (1.6) at time $t$ in terms of the dressed exciton and phonon operators

$$\hat{a}_n(t) = e^{\frac{i}{\hbar}\hat{H}_{MF}t} \hat{A}_n e^{\sum_q g_q e^{iqn}(\hat{B}_q - \hat{B}_{-q}^\dagger)} e^{-\frac{i}{\hbar}\hat{H}_{MF}t} = \hat{A}_n e^{-i\Delta t} e^{\sum_q g_{qn} e^{iqn}(\hat{B}_q e^{-i\omega_q t} - \hat{B}_{-q}^\dagger e^{i\omega_{-q} t})}. \qquad (1.9)$$

Power spectrum for the spontaneous photon emission power can be defined as the Fourier transform

$$S(\omega) = \text{Re} \int_0^\infty dt\, e^{i\omega t} C_{k=0}(t) \qquad (1.10)$$

of the exciton auto-correlation function

$$C_k(t) = \langle \hat{a}_k^\dagger(0) \hat{a}_k(t) \rangle \qquad (1.11)$$

represented in the momentum space. By setting the exciton momentum $k = 0$, we account for the negligibly small momentum of the emitted photons. The angular breakers in Eq. (1.10) denote the average with the reduced density operator for the excitons and phonons. The time-evolution of the exciton operator is governed by the Hamiltonian (1.8). By performing the Fourier transformation of Eq. (1.11), we find the desired time dependance for the bare exciton operator

$$\hat{a}_k(t) = \sum_n \hat{A}_n e^{ikn} \, e^{\sum_q g_q \left( \hat{B}_q e^{-i\omega_q t} - \hat{B}^\dagger_{-q} e^{i\omega_{-q}t} \right)} e^{iqn}. \tag{1.12}$$

Making substitution of Eq. (1.12) and its Hermitian conjugate at time zero into Eq. (1.11), we find the exciton auto-correlation function

$$C_k(t) =$$
$$\sum_{nn'} \langle \hat{A}^\dagger_n \hat{A}_{n'} \rangle e^{-i\Delta t + ik(n-n')} \left\langle 0 \left| e^{-\sum_q g_q^* \left( \hat{B}_{-q} - \hat{B}^\dagger_q \right)} e^{-iqn} \, e^{\sum_q g_q \left( \hat{B}_q e^{-i\omega_q t} - \hat{B}^\dagger_{-q} e^{i\omega_{-q}t} \right)} e^{iqn'} \right| 0 \right\rangle, \tag{1.13}$$

where we used a reasonable assumption that at zero time the reduced density operator partitions into the product of the phonon vacuum density $|0\rangle\langle 0|$ and the exciton density matrix $\langle \hat{A}^\dagger_n \hat{A}_{n'} \rangle$. Further applying the Baker-Hausdorff formula $e^{\hat{F}+\hat{G}} = e^{\frac{1}{2}[\hat{F},\hat{G}]} e^{\hat{G}} e^{\hat{F}}$ and the relationship $g^*_q = g_{-q}$, we evaluate the phonon matrix element in Eq. (1.13)

$$\left\langle 0 \left| e^{-\sum_q g_q^* \left( \hat{B}_{-q} - \hat{B}^\dagger_q \right) e^{-iqn}} \, e^{\sum_q g_q \left( \hat{B}_q e^{-i\omega_q t} - \hat{B}^\dagger_{-q} e^{i\omega_{-q}t} \right) e^{iqn'}} \right| 0 \right\rangle$$

$$= e^{i\sum_q |g_q|^2 \sin(\omega_q t + q(n-n'))}$$

$$\times \left\langle 0 \left| e^{\sum_q g_q \hat{B}_q \left( e^{-i\omega_q t + iqn'} - e^{iqn} \right) - \sum_q g_q \hat{B}^\dagger_q \left( e^{i\omega_q t - iqn'} - e^{-iqn} \right)} \right| 0 \right\rangle$$

$$= e^{\sum_q |g_q|^2 \{i \sin(\omega_q t + q(n-n')) + \cos(\omega_q t + q(n-n')) - 1\}}$$

$$\times \left\langle 0 \left| e^{-\sum_q g_q \hat{B}^\dagger_q \left( e^{i\omega_q t - iqn'} - e^{-iqn} \right)} e^{\sum_q g_q \hat{B}_q \left( e^{-i\omega_q t + iqn'} - e^{iqn} \right)} \right| 0 \right\rangle$$

$$= e^{\sum_q |g_q|^2 \left( e^{i\omega_q t + iq(n-n')} - 1 \right)}. \tag{1.14}$$

This calculation simplifies the exciton auto-correlation function (1.14), to the form

$$C_k(t) = e^{-\sum_q |g_q|^2} \sum_{nn'} \langle \hat{A}^\dagger_n \hat{A}_{n'} \rangle e^{-i\Delta t + ik(n-n')} e^{\sum_q |g_q|^2 e^{i\omega_q t + iq(n-n')}}. \tag{1.15}$$

A final touch in our derivation is to express the exciton density matrix via the exciton Wigner distribution function

$$\langle \hat{A}^\dagger_n \hat{A}_{n'} \rangle = \sum_\kappa W\left( \frac{n+n'}{2}, \kappa \right) e^{i\kappa(n'-n)}, \tag{1.16}$$

of momentum $\kappa$ and coordinate $(n+n')/2$. Introducing this result in Eq. (1.15), we obtain

$$C_k(t) = e^{-\sum_q |g_q|^2} \sum_{n\kappa} f_\kappa \, e^{-i\Delta t + i(k-\kappa)n} e^{\sum_q |g_q|^2 e^{i\omega_q t + iqn}}, \tag{1.17}$$

as a function of the momentum space exciton distribution function $f_\kappa \equiv \sum_{n+n'} W\left( \frac{n+n'}{2}, \kappa \right)$.

To complete the calculation, we expand the last exponential in Eq. (1.17) into the Taylor series and perform the summation over $n$ and $\kappa$ to obtain

$$C_k(t) = e^{-\sum_q |g_q|^2} \sum_{m=0}^\infty \sum_{q_1} \cdots \sum_{q_m} \frac{f_{k+q_1+\cdots+q_m}}{m!} |g_{q_1}|^2 \cdots |g_{q_m}|^2 \, e^{-i(\Delta - (\omega_{q_1} + \cdots + \omega_{q_m}))t}. \tag{1.18}$$

In this expression, we sum over all possible $m$-phonon scattering events contributing to the exciton auto-correlation function. It is important to notice that the distribution function for the exciton momenta, $f_{k+q_1+\cdots+q_m}$, accounts for the phonon recoil effect as required by the momentum conservation, i.e., explicitly depends on $k + q_1 + \cdots + q_m$. Therefore, $f_{k+q_1+\cdots+q_m}$ is an important parameter defining relative contribution of each multi-phonon process.

Making the substitution of Eq. (1.18) into Eq. (1.10) and evaluating the Fourier integral, we find desired expression for the photon emission lineshape

$$S(\omega) = e^{-\Sigma_q |g_q|^2} \sum_{m=0}^{\infty} \sum_{q_1} \cdots \sum_{q_m} \frac{f_{q_1+\cdots+q_m}}{m!} |g_{q_1}|^2 \cdots |g_{q_m}|^2 \frac{\gamma/\pi}{(\omega-\Delta+\omega_{q_1}+\cdots+\omega_{q_m})^2+\gamma^2}, \quad (1.19)$$

where the exciton dephasing rate $\gamma$ is introduced. Notice that in accords with the discussion above, the exciton momentum distribution function $f_{q_1+\cdots+q_m}$ defines spectral weight of each m-photon spectral replica. Eq. (1.19) significantly simplifies under the assumption that the exciton-phonon coupling, $g_q$, and optical phonon frequency, $\omega_q$, have flat dispersion curve, i.e., independent of the momentum $q$. Respectively, denoting them as $g$ and $\omega_{ph}$ we recast Eq. (1.19) to the form

$$S(\omega) = e^{-N|g|^2} \sum_{m=0}^{\infty} \frac{|g|^{2m}}{m!} \frac{A_m \gamma/\pi}{(\omega-\Delta+m\omega_{ph})^2+\gamma^2}, \quad (1.20)$$

where $N$ is the total number of sites in the lattice. In Eq. (1.20) each m-phonon lineshape function is weighted with the factor

$$A_m = \sum_{q_1} \cdots \sum_{q_m} f_{q_1+\cdots+q_m}. \quad (1.21)$$

A simple estimate for $A_m$ can be performed assuming that the exciton momentum distribution is sharply peaked around zero, i.e., the exciton is spatially delocalized over all $N$ sites of the lattice. In this case the distribution function can be approximated as $f_q = \left(\frac{2\pi}{a}\right)\delta(q)$, where $a$ is the lattice period. Performing intergradation over the Brillouin zone in Eq. (1.21), we find that $A_m = N^m$. This result shows that the exciton-phonon coupling scales as $\sqrt{N}g$ allowing for the thermodynamic limit making application of our theory scalable finite size nano-structures to the bulk limits.

**Fitting PL spectra using exciton-polaron Holstein model**

To fit the experimental data, we generalize Eq. (1.21) with $A_m = N^m$ to the case of three independent vibration modes. The result is

$$S(\omega) = S_o e^{-g_1^2-g_2^2-g_3^2} \sum_{m_1=0}^{M_{max}} \sum_{m_2=0}^{M_{max}} \sum_{m_3=0}^{M_{max}} \frac{g_1^{2m_1} g_2^{2m_2} g_3^{2m_3}}{m_1! m_2! m_3!} \times$$
$$\frac{(m_1 \gamma_{ph1}+m_2 \gamma_{ph2}+m_3 \gamma_{ph3}+\gamma)}{(\omega-\Delta+m_1\omega_{ph1}+m_2\omega_{ph2}+m_3\omega_{ph3})^2+(m_1\gamma_{ph1}+m_2\gamma_{ph2}+m_3\gamma_{ph3}+\gamma)^2}, \quad (1.23)$$

where for the sake of brevity we replaced $\sqrt{N}|g| \to g$ for the exciton-phonon coupling parameters and added the phonon line broadening $\gamma_{ph}$.

Results of the data fit with the three-mode model (1.23) are presented in Fig. SI-8 and Table SI-1. The model reproduces well the main features of the experimental plot. Extracted values for the exciton-phonon coupling parameter, $g$, are all less than one suggesting relatively weak coupling regime which is consistent with the observed Poissonian as opposed to the Gaussian type (blue and gray lines) of the phonon replica progression.

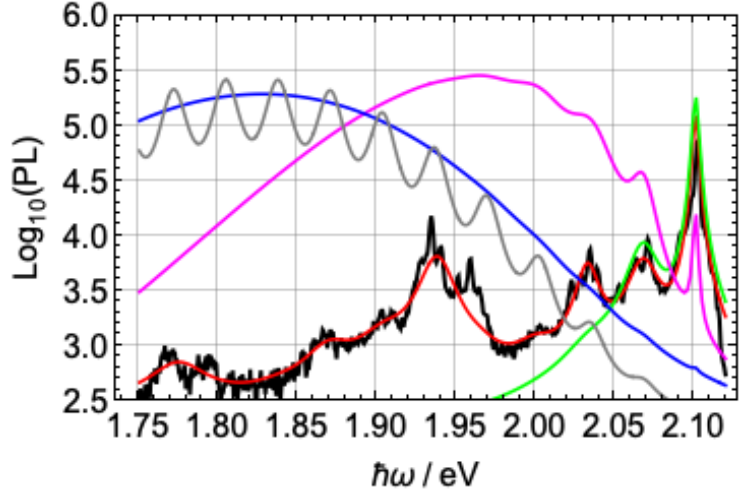

**Fig. SI-8: Holstein exciton-polaron model fits.** Log plot of experimental PL spectra (solid black) from Fig. 3a in the main text overlaid with line-shape fits (red line) using Eq. (1.23) derived using the adopted Holstein exciton-polaron model. Fit parameters are summarized in Table SI-1. Using the fit parameter for Phonon mode 1 (Tables SI-1), we calculated PL spectrum (green line) for the single phonon mode model. An increase of the exciton phonon coupling to $g_1 = 2.3$ and $g_1 = \sqrt{9.9} = 3.15$ in this model results in the PL lineshapes shown by the magenta and blue lines, respectively. Further changing the line broadening parameters to $\hbar\gamma_{ph} = 0$ meV and $\hbar\gamma = 10$ meV in the single mode calculation but retaining $g_1 = \sqrt{9.9} = 3.15$ allowed us to resolve the phonon mode replica (gray line) washed out in the blue line. Note: the gray line resembles the spectrum shown by Ergecen, E. et al. in *Nat Commun* **13**, 98, (2022) where the strong exciton-phonon coupling results in washing out of the zero-phonon line.

**Table SI-1: Fit parameters.** Parameters obtained from fitting PL shown on Fig. SI-8 using Eq. (1.23) with $M_{max} = 10$. Zero phonon line is characterized by $\hbar\Delta = 2.1 \pm 2 \times 10^{-4}$ eV and $\hbar\gamma = 2.4 \pm 0.2$ meV. Obtained normalization perfector is $S_o = 1293 \pm 36$.

|  | $\hbar\omega_{ph}$ / meV | $\hbar\gamma_{ph}$ / meV | $g$ |
|---|---|---|---|
| **Phonon mode 1** | 32.8 ± 0.7 | 7.4 ± 1.3 | 0.49 ± 0.03 |
| **Phonon mode 2** | 68.5 ± 0.7 | 4.8 ± 1.0 | 0.39 ± 0.02 |
| **Phonon mode 3** | 163.0 ± 0.6 | 8.8 ± 0.9 | 0.58 ± 0.01 |

## S.6. Charging Model for Spectral Diffusion

The model has two essential parts. First, the inhomogeneous broadening of the fluorescence spectrum is assumed to be caused by fluctuations in the exciton binding energy. These fluctuations are proportional to fluctuations in the intralayer electric field through the first-order (linear) Stark shift. Second, electric field fluctuations are caused by sudden shifts of the charge density in the nanoflake, which arise from interlayer charge transfer due to hopping of hot electrons and holes just after laser excitation and before thermal relaxation.

**Linear Stark Shift**

Let us assume that the stochastic jumping of the exciton emission spectrum, with periodic frequency shifts in multiples of ~1 meV, is due to fluctuations in the Stark shift of the excitons confined to two-dimensional layers due to charge transfer between adjacent layers. In first order perturbation, the energy shift

$$\delta = -\vec{\mu} \cdot \vec{E}, \quad (2.1)$$

associated with exciton emission due to an electric field $E$ depends on the electric dipole operator $\mu$. For an electric dipole moment $\mu \sim 10D$, a shift of 1 meV requires fields as high as $10^5$ V/cm.

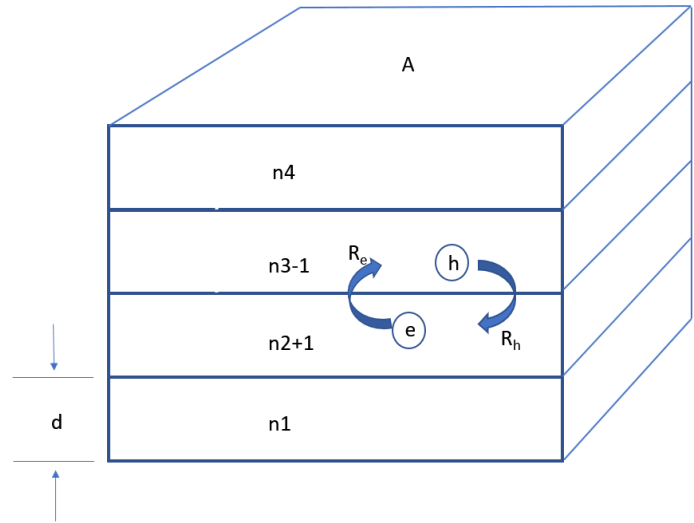

**Fig. SI-9: Parallel capacitor model.** Shown here is schematic for a 4-layer nanoflake, modeled as a set of 4 parallel conducting layers with area A and layer periodicity d. The $i^{th}$ layer has charge $q_i = n_i e$ which is an integral multiple of the electronic charge, with positive and negative integer $n_i$ corresponding to whether the excess charge is comprised of holes or electrons, respectively. A change in state occurs when either the hole or the electron of an energetic electron-hole pair, created by the laser in any one of the 4 layers, hops to an adjacent layer before complete thermal relaxation can occur. In this illustration, we assume that the transfer occurs between layers 2 and 3, with $n_2 \rightarrow n_2 + 1$, and $n_3 \rightarrow n_3 - 1$. This can happen with the transfer of an electron from layer 2 to layer 3, or with the transfer of a hole from layer 3 to layer 2.

To first approximation, consider the nanoflake to act as a capacitor, with each layer serving as a uniformly charged plate. In such a case, the electric field,

$$E = \frac{e}{\kappa \varepsilon_0 A} \sim 10^5 \text{V/cm}, \tag{2.2}$$

due to single-electron charging is the right order of magnitude for a 1 meV Stark shift if the product $\kappa A \simeq 2 \times 10^3 \text{nm}^2$. For a dielectric constant $\kappa \simeq 5$ in agreement with calculations for bulk NiPS$_3$,[1] this corresponds to an area $A \simeq 400 \text{nm}^2$ suggesting lateral dimensions on the order of 20 nm. This is within reasonable agreement of AFM measurements reported in Fig. 1b.

**Charge Hopping Rates**

Introduction of a rate model for the hopping of electrons and holes sharpens this discussion. Consider a nanoflake consisting of $N$ parallel conducting layers, each with an area $A$, separated from one another with a thickness $d$. Fig. SI-9 shows a schematic of a four-layer structure as an illustration. The $i^{\text{th}}$ layer has charge $q_i = n_i e$, which is an integral multiple of the electronic charge, with positive and negative integer $n_i$ corresponding to whether excess charge is comprised of holes or electrons, respectively. Thus, the charge state of the nanoflake is determined by the set of integers $\{n_i\}$. A change in state occurs when either the hole or the electron of a laser-produced energetic electron-hole pair, created in any one of the layers, hops to an adjacent layer before thermal relaxation can occur. Fig. SI-9 illustrates the case where charge transfer occurs between layers 2 and 3, with $n_2 \rightarrow n_2 + 1$, and $n_3 \rightarrow n_3 - 1$. The rate

$$R_{f,i} = R \begin{cases} \exp\left(-\left(U_f - U_i\right)/kT\right) & ; U_f > U_i \\ 1 & ; U_f < U_i \end{cases} \tag{2.3}$$

for changing from an initial state $\{n_1, n_2, n_3, n_4\}_i$ to a final state $\{n_1, n_2+1, n_3-1, n_4\}_f$ is given by the sum $R = R_e + R_h$ of electron and hole hopping rates. The Boltzmann factor $\exp\left(-\left(U_f - U_i\right)/kT\right)$ expresses detailed balance between initial and final states with energies $U_i$ and $U_f$ respectively. After excitation, an electron hole pair will immediately begin thermalizing to a temperature of 10 K (~1 meV), after which hops between layers are frozen out. Thus $kT$ in Eq. (2.3) is an effective thermal energy for rare charge-transfer events, expected to lie somewhere between 1 meV and 200 meV, the difference $h\nu - \varepsilon_g$ between the 2.4 eV laser excitation energy and the 2.2 eV band gap. The initial and final charge-state energies can be determined by the spatial integration of the electrical energy density $\frac{1}{2}\varepsilon E^2$. If the charge is confined to thin conducting planes in each layer, the energy

$$U(\{n_i\}) = \frac{1}{2}\kappa\varepsilon_0 A d \sum_{j=1}^{N-1} E_j^2 = \frac{e^2}{2C} \sum_{j=1}^{N-1} \left(\sum_{i=1}^{j} n_i\right)^2, \tag{2.4}$$

can be written as a sum over the square of the electric field penetrating the potential-barrier regions separating each of the planes. For a nanoflake with $N$ layers, there are $N$-1 such regions. In writing the second equality in Eq. (2.4) we have used Gauss's law to express the electric field $E_j = e\sum_{i=1}^{j} n_i / \kappa\varepsilon_0 A$ in the $j^{th}$ barrier in terms of the charge $e\sum_{i=1}^{j} n_i$ distributed in the conducting layers preceding it. We have also introduced the capacitance of a single layer,

$$C = \frac{\kappa\varepsilon_0 A}{d} \simeq 4 \times 10^{-17} F , \qquad (2.5)$$

for a typical interlayer distance $d \sim 0.5$nm. Thus, the energy change between initial and final states will be quantized in multiples of $e^2/2\,C \simeq 2$meV.

**Kinetic Monte Carlo**

Spectral diffusion can be simulated by following the charge state of the nanoflake in time in a Monte Carlo simulation of a continuous-time random walk.[2] For an *N*-layer system, a given charge state,

$$\{n_1, n_2, n_3, n_4, \ldots n_N\} \rightarrow \begin{array}{l} \{n_1 \pm 1, n_2 \mp 1, n_3, n_4, \ldots n_{N-1}, n_N\} \\ \{n_1, n_2 \pm 1, n_3 \mp 1, n_4, \ldots n_{N-1}, n_N\} \\ \{n_1, n_2, n_3 \pm 1, n_4 \mp 1, \ldots n_{N-1}, n_N\} \\ \ldots \\ \{n_1, n_2, n_3, n_4, \ldots n_{N-1} \pm 1, n_N \mp 1\} \end{array} \qquad (2.6)$$

is connected to 2(*N*-1) neighboring states enumerated here by hopping rates $R_{f,i}$ of the form given in Eq. (2.3). For each step of the random walk, the initial energy $U_i(\{n\})$ associated with presently occupied state, and the final energy $U_f(\{n\})$ for each of the neighbors, are used to calculate 2(*N*-1) different values of the hopping rate Eq. (2.3). The final state is then determined from a uniform distribution according to its relative weight, $w_{f,i} = R_{f,i} / \sum_{f'} R_{f,i}$. At the same time, a dwell time associated with the initial state is selected from an exponential distribution of dwell times,

$$\psi_i(t) = \frac{1}{\tau_i} e^{-t/\tau_i} , \qquad (2.7)$$

where the lifetime $\tau_i = \left(\sum_f R_{f,i}\right)^{-1}$ is the reciprocal of the sum of all 2(*N*-1) rates. In Fig. SI-10, we follow the charging effect on the Stark shift in an *N*-layer system in time for a single random walk, to qualitatively compare results of the simulation to the observed temporal evolution of the spectral jumping. We also make a time-weighted histogram of results, Fig. SI-11, to determine the steady-state probability for an exciton to emit with a particular Stark shift.

**Electric Field within a Layer**

Explicit calculations of the Stark shift (Fig. SI-10 and Fig. SI-11) require knowledge of average electric field applied to the exciton states in each layer. An exciton embedded in a 2-D charged plane is subjected to the average,

$$E_{j(\text{layer})} = \frac{e}{2\kappa\varepsilon_0 A}\left(\sum_{i=1}^{j-1} n_i + \sum_{i=1}^{j} n_i\right) = \frac{e}{2\kappa\varepsilon_0 A}\left(2\sum_{i=1}^{j} n_i - n_j\right), \qquad (2.8)$$

of the electric field immediately above and below the plane. According to (2.8), if there is no charge in the $j^{\text{th}}$ layer, i.e., if $n_j = 0$, then the field in the $j^{\text{th}}$ layer is an even multiple of $e/2\kappa\varepsilon_0 A$. If $n_j \neq 0$, then the field in the $j$-th layer will be an even multiple of $e/2\kappa\varepsilon_0 A$ for even $n_j$, and an odd multiple of $e/2\kappa\varepsilon_0 A$ for odd $n_j$. If charge residing within a layer quenches exciton emission such that only uncharged layers participate in the spectrum, the Stokes shift will be quantized in multiples of $e\mu/\kappa\varepsilon_0 A \sim 2\text{meV}$. If charge residing with a layer has no quenching effect, then the Stark shift will be quantized in multiples of $e\mu/2\kappa\varepsilon_0 A \sim 1\text{meV}$. Partial quenching gives a superposition of the two intervals, in agreement with observation. For the plots in Fig. SI-10 and Fig. SI-11, we have introduced partial quenching by allowing layers with no charge and/or with one excess charge to be four times more likely to emit photons than any of the other charge states.

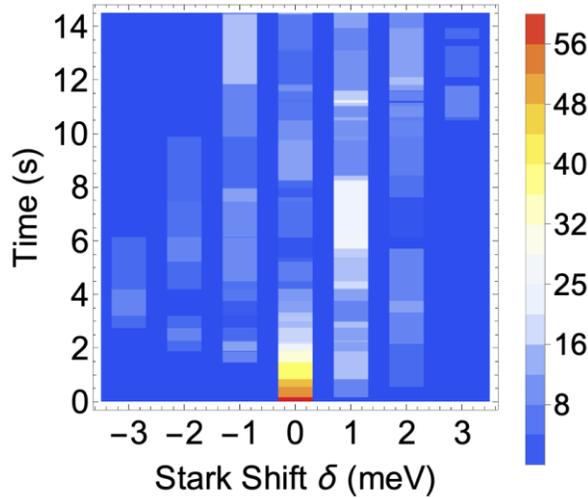

**Fig. SI-10**: **Spectral diffusion.** Shown is an illustration of spectral diffusion versus time, following the evolution of the continuous time random walk for 50 hops for a nanoflake with 15 layers. The horizontal axis displays the Stark shift $\delta$, and time is displayed on the vertical axis. The intensity of emission is indicated by the color temperature on a scale from 0 (blue) to a maximum of 60 (red). Abrupt changes in color indicate hops. Initially $\delta = 0$, none of the layers has yet to acquire excess charge, and the intensity maximum. After 50 hops, $\delta$ ranges from -3 meV to 3 meV. The dispersion in $\delta$ is controlled by a single dimensionless parameter $(e^2/2C)/kT = 0.02$. For $e^2/2C = 2\text{meV}$ this corresponds to $kT \simeq 0.1\text{eV}$. The Stark shift is quantized in multiples of $\mu e/2\kappa\varepsilon_0 A = 1\text{meV}$, while the length of the time-axis determined by the hopping rate. Here $R = 0.1\text{s}^{-1}$.

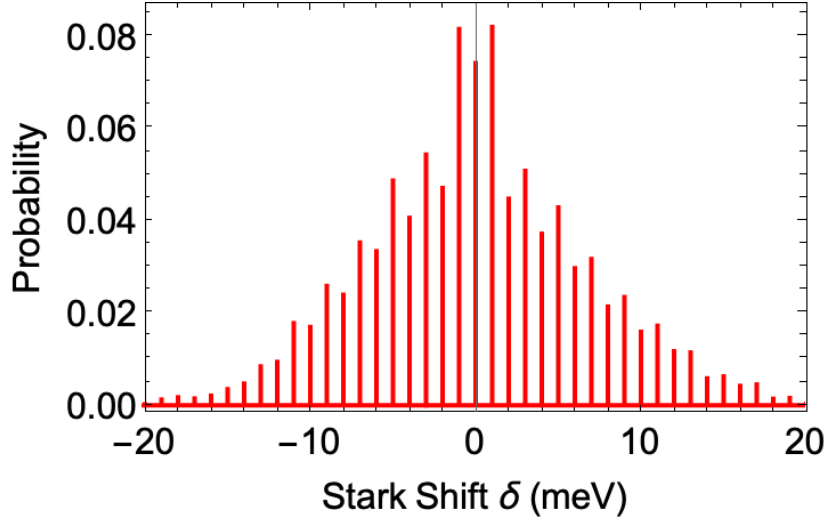

**Fig. SI-11: Stark shift comb.** The figure shows the steady-state probability to observe different values of $\delta$ for a 15-layer nanoflake. The same parameter values were used for Fig. SI-10. The histogram displays the time-weighted spectrum for 20,000 hops. The spectrum shifts in multiples of 1.0 meV. We have taken the probability for radiative decay to be 4 times more likely in layers with no charge, $n = 0$, and/or one charge, $n = \pm 1$, and this weighting gives rise to the alternating pattern shown here. Correlation between layers having few charges and layers having low electric fields causes an enhanced probability in the neighborhood of $\delta \simeq 0$.

**Fine Structure of PL Spectrum**

The simplest model for the emission spectrum is to assume linear exciton-phonon coupling to a single dispersionless optical phonon band with frequency $\omega_0$.[3] Neglecting gauge terms generated in the Lang-Firsov transformation, the low-temperature spectrum (follows from the Holstein Model in SI-1 with infinitesimal line widths),

$$S(\omega) = e^{-g^2} \sum_{m=0}^{\infty} \frac{g^{2m}}{m!} \delta(\omega - \Delta + m\omega_0), \tag{2.9}$$

is a set of equally spaced lines, beginning with the 0-0 line at $\omega = \Delta$, with successively lower energy phonons in multiples of $\omega_0$ weighted by a Poissonian distribution determined by the square of the dimensionless coupling constant $g = \delta x / \sqrt{2} x_0$ measuring the oscillator displacement $\delta x$ in multiples of the zero point amplitude $x_0$.[3] Let us assume that the exciton is linearly coupled to three optical phonons, with energies $\hbar\omega_1 = 31\,\text{meV}$, $\hbar\omega_2 = 68\,\text{meV}$, and $\hbar\omega_3 = 168\,\text{meV}$. (The phonon energies are slightly tweaked from the values obtained from the low-resolution PL spectra fit in Table SI-1.) A line spectrum can be constructed that reproduces most of the experimental features in Fig. 3a (except for the P4 line) by choosing coupling constants $g_1 = 0.49$, $g_2 = 0.38$, and $g_3 = 0.58$, respectively. The Stark shift comb described above and shown in Fig. SI-10 has been convoluted to provide inhomogeneous broadening. The smoothed histogram was made with Gaussian bins having standard deviation 0.35 meV. The result of the calculation is shown in Fig. SI-12 (red line) demonstrating reasonable agreement with the experimental data (black line).

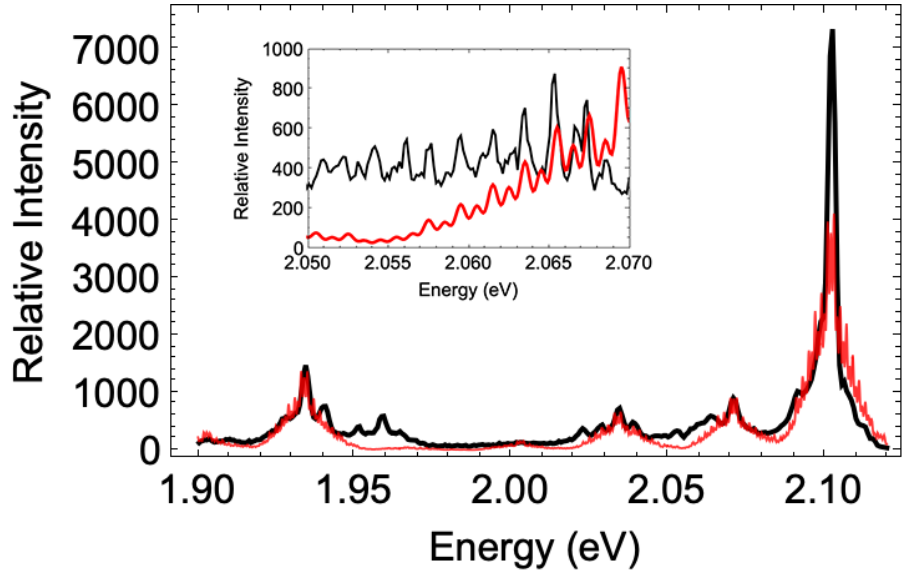

**Fig. SI-12: Spectrum with three commensurate phonons.** Shown is a calculated spectrum in which the exciton is weakly coupled to three different phonons, with frequencies that are approximately multiples of 33 meV, as described in the text. This has been superimposed on the measured spectrum shown in Fig. 2a, with alignment of the 0-0 lines at 2.102 eV. The only source of broadening in the calculated spectrum is the inhomogeneous broadening provided by the different electrical environments in the different layers, and this has been incorporated by convoluting the line spectra with the Stark shift comb shown in Fig. SI-11. A magnification (inset) of the calculated spectrum (red) reveals the periodicity of the comb. The energy spacing between high peaks is 2 meV, while the energy spacing from low peak to high peak 1 meV. The alternating structure is similar to that seen in the measured spectrum at high resolution (black).

**References:**

1	Lane, C. & Zhu, J.-X. Thickness dependence of electronic structure and optical properties of a correlated van der Waals antiferromagnetic NiPS3 thin film. *Phys. Rev. B* **102**, 075124 (2020).
2	Young, W. M. & Elcock, E. W. Monte Carlo studies of vacancy migration in binary ordered alloys: I. *Proc. Phys. Soc.* **89**, 735-746 (1966).
3	Huang, K., Rhys, A. & Mott, N. F. Theory of light absorption and non-radiative transitions in F-centres. *Proc. R. Soc. Lond. A* **204**, 406-423 (1950).